\documentstyle[12pt]{article}
\begin{document}
\begin{center}
{\Large On de Sitter deflationary cosmology from the spin-torsion primordial fluctuations and COBE data}
\vspace{1cm}
\noindent

L.C. Garcia de Andrade\footnote{Departamento de Fisica Teorica,Instituto de F\'{\i}sica , UERJ, Rua S\~{a}o
francisco Xavier 524, Rio de Janeiro,CEP:20550-013, Brasil.e-mail:garcia@dft.if.uerj.br.}
\end{center}
\vspace{2cm}
\begin{center}
{\Large Abstract}
\end{center}
\vspace{0.5cm}
Fluctuations on de Sitter solution of Einstein-Cartan field equations are obtained in terms of the matter density primordial density fluctuations and spin-torsion density and matter density fluctuations obtained from COBE data.
Einstein-de Sitter solution is shown to be unstable even in the absence of torsion.The spin-torsion density fluctuation to generate a deflationary phase is computed from the COBE data. 
\newpage
Recently D.Palle \cite{1} has computed the primordial matter density fluctuations  from COBE satellite data.More recently I have been extended Palle's work to include the dilaton fields \cite{2}.Also more recently I have considered a mixed inflation model with spin-driven inflation and inflaton fields \cite{3} where the spin-torsion density  have been obtained from the COBE data on the temperature fluctuations.However in these last attempts no consideration was given on the spinning fluid and torsion was considered just coming from the density of inflaton fields which of course could be considered just in the case of massless neutrinos.Earlier Maroto and Shapiro \cite{4} have discussed the de Sitter metric fluctuations and showed that in the case with dilatons and torsion the higher-order gravity the stability of de Sitter solutions depends on the parametrization and dimension,but that for the given dimension one can always choose parametrization in such a way that the solutions are unstable.In this letter we show that starting from the Einstein-Cartan equations as given in Gasperini for a four-dimension spacetime with spin-torsion density the de Sitter solutions are also unstable for large values of time.Of course one should remind that Maroto-Shapiro solutions  does not possess spin but are based on a string type higher order gravity where torsion enters like in Ramond action.Let us start from the Gasperini \cite{5} form of the Einstein-Cartan equations for the spin-torsion density
\begin{equation}
H^{2}=\frac{8{\pi}G}{3}({\rho}-2{\pi}G{\sigma}^{2})
\label{1}
\end{equation}
and
\begin{equation} 
{\dot{H}}+H^{2}=-\frac{4{\pi}G}{3}({\rho}+3p-8{\pi}G{\sigma}^{2})  
\label{2} 
\end{equation}
where $\frac{\ddot{a}}{a}=H(t)$ where $H(t)$ is the Hubble parameter.Following the linear perturbation method in cosmological models as described in Peebles \cite{6} we have 
\begin{equation}
H(t,r)=H(t)[1+{\alpha}(r,t)]
\label{3}
\end{equation}
where ${\alpha}=\frac{{\delta}H}{H}$ is the de Sitter metric density fluctuations where the Friedmann metric is reads
\begin{equation}
ds^{2}=dt^{2}-a^{2}(dx^{2}+dy^{2}+dz^{2})
\label{4}
\end{equation}
Also the matter density is given by
\begin{equation}
{\rho}(r,t)={\rho}(t)[1+{\beta}(r,t)]
\label{5}
\end{equation}
and 
\begin{equation}
{\sigma}(r,t)={\sigma}(t)[1+{\gamma}(r,t)]
\label{6}
\end{equation}
whhere ${\beta}=\frac{{\delta}{\rho}}{{\rho}}$ is the matter density fluctuation which is approximately $10^{-5}$ as given by COBE data and ${\gamma}(r,t)=\frac{{\delta}{\sigma}}{{\sigma}}$ where ${\sigma}$ is the spin-torsion density.Substitution of these equations into the Gasperini-Einstein -Cartan equations above in the simple case where the pressure $p$ vanishes (dust) we obtain 
\begin{equation}
\dot{\alpha}=\frac{\dot{H}}{H}+\frac{8{\pi}G}{3H}[{\rho}{\beta}-16{\pi}G{\sigma}{\gamma}]
\label{7}
\end{equation}
This last equation can be integrated to 
\begin{equation}
{\alpha}=1+ln{H}+{8{\pi}G}[\int{\frac{{\rho}{\beta}dt}{H}}-16{\pi}G\int{\frac{{\sigma}{\gamma}dt}{H}}]
\label{8}
\end{equation}
When the variation of mass density and spin density are small with respect to time they can be taken as approximatly constant and we are able to perform the integration for de Sitter metric $(H=H_{0}=constant)$ as
\begin{equation}
{\alpha}=1+ln{H_{0}}+\frac{8{\pi}G}{3H_{0}}t[{{\rho}_{0}}{\beta}-16{\pi}G{{\sigma}_{0}}{\gamma}]
\label{9}
\end{equation}
which shows clearly that the de Sitter solution to Einstein-Cartan equations is unstable and can be computed in terms of the matter and spin-torsion densities.Let us now compute the spin-torsion fluctuation from expression (\ref{9}) necessary to produce deflation $(\frac{\dot{a}}{a}<0)$.By making use of the mean cosmological matter density ${\rho}_{0}=10^{-31}g.cm^{-3}$,the spin-torsion density computed in (\ref{7}) ${\sigma}_{0}=10^{18}$ and the COBE data for the matter density fluctuation in the Universe along with the condition of small density perturbations ${\alpha}<<1$ in formula (\ref{9}) after a straightforward algebra one obtains
\begin{equation}
\frac{{\delta}{\sigma}}{{\sigma}}=10^{-54}cgs-units
\label{10}
\end{equation}
This result shows that a slight fluctuation in the spin-torsion density maybe able to produce a deflationary phase in the de Sitter Universe.If no deflationary phases are not found in the universe it may happen the torsion is not necessary to describe spin and this model is wrong or that deflationary phase have been not yet observed in the Universe.
One may rewrite expression (\ref{2}) as
\begin{equation}
(\frac{\ddot{a}}{a}<0)=\frac{\dot{a}}{{a}^{2}}-\frac{4{\pi}G}{3}({\rho}-8{\pi}G{\sigma}^{2})
\label{11}
\end{equation}
Since the other deflation condition is $(\dot{a}<0)$ one obtains from equation (\ref{11}) a constraint on the matter density given by
\begin{equation}
{\rho}<8{\pi}G{\sigma}^{2}
\label{12}
\end{equation}
Which clearly shows that the de Sitter deflation occurs on a spin-torsion dominated phase of the Universe.This situation may lead to violations which may lead to the presence of wormholes or even dark matter. 
\vspace{2cm}
\begin{flushleft}
{\large Acknowledgements}
\end{flushleft}
I would like to thank Professor Ilya Shapiro and Prof.Rudnei Ramos for helpful discussions on the subject of this paper.Financial support from CNPq. is gratefully ackowledged.

\end{document}